\documentclass[11pt]{article}
\usepackage[utf8]{inputenc}
\usepackage[top=1in, bottom=1in, left=1in, right=1in]{geometry}
\usepackage{graphicx,tikz,enumitem,framed}
\usepackage[hyperindex,breaklinks]{hyperref}
\usepackage{booktabs,soul,color}
\usepackage{rotating}
\usepackage[flushleft]{threeparttable}
\usepackage{amsmath, amsfonts, amssymb,hhline,caption}
\hypersetup{
	breaklinks=true,   
}

\renewcommand{\vec}[1]{{\boldsymbol{#1}}}

\newcommand{\ddelta}{\vec{\delta}}	
\newcommand{\eeta}{\vec{\eta}}
\newcommand{\bbeta}{\vec{\beta}}
\newcommand{\aalpha}{\vec{\alpha}}	

\newcommand{\Cov}{\text{cov}}

\newcommand{\EEalpha}{U_i^m(\aalpha,\ddelta, \eeta)}

\renewcommand{\P}{\text{P}}

\newtheorem{Corollary}{Corollary}

\title{Adjusting for Participation Bias in Case-Control Genetic Association Studies for Rare Diseases}
\author{
	Le Wang\footnote{Corresponding author: \texttt{le.wang@lmu.edu}}\ \,\footnote{Two authors contributed equally to the method development.}\ \,\thanks{Department of Mathematics, Statistics and Data Science, Loyola Marymount University, Los Angeles, California, 90045, USA}
	\and  
	Zhengbang Li\footnotemark[2]\ \,\thanks{School of Mathematics and Statistics, Central China Normal University,  Wuhan, Hubei, 430079, PRC}
	\and 
	Ben Fitzpatrick\footnotemark[3]
	\and 
	Clarice Weinberg\thanks{Biostatistics and Computational Biology Branch, National Institute of Environmental Health Sciences, Durham, North Carolina, 27709, USA}
	\and
	Jinbo Chen\thanks{Department of Biostatistics, Epidemiology and Informatics, University of Pennsylvania, Philadelphia, Pennsylvania, 19104, USA}
}

\date{
}

\begin{document}
	\maketitle
	
\begin{abstract}
Collection of genotype data in case-control genetic association studies may often be incomplete for reasons related to genes themselves. This non-ignorable missingness structure, if not appropriately accounted for, can result in participation bias in association analyses. To deal with this issue, Chen et al. \cite{chen2016} proposed to collect additional genetic information from family members of individuals whose genotype data were not available, and developed a maximum likelihood method for bias correction. In this study, we develop an estimating equation approach to analyzing data collected from this design that allows adjustment of covariates. It jointly estimates odds ratio parameters for genetic association and missingness, where a logistic regression model is used to relate missingness to genotype and other covariates. Our method allows correlation between genotype and covariates while using genetic information from family members to provide information on the missing genotype data. In the estimating equation for genetic association parameters, we weight the contribution of each genotyped subject to the empirical likelihood score function by the inverse probability that the genotype data are available. We evaluate large and finite sample performance of our method via simulation studies and apply it to a family-based case-control study of breast cancer. 

\bigskip
\noindent
{\bf Keywords:} Participation bias, Empirical likelihood, Inverse-probability-weighting, Non-ignorable missingness, Genetic association, Case-control studies.
\end{abstract}

\section{Introduction}
\label{intro}
In case-control genetic association studies, genotype data may be missing for reasons related to the phenotype under study. Such missingness, if ignored, may lead to biased association analyses when the genetic variant of interest is associated with both the phenotype and missingness. Such non-ignorable missingness is commonly referred to as the ``participation bias". For highly lethal diseases, genotype data of deceased cases may be selectively excluded from the study due to disease-specific and possibly gene-related mortality, or a patient's advanced disease stage may prevent genotype data collection or informed consent. Participation bias has been reported in genetic association studies of age-related macular degeneration \cite{chiu2011}, ovarian cancer  \cite{lacour2011}, coronary heart disease \cite{williams2011}, and other cardiovascular diseases \cite{anderson2011,falcone2013,horsfall2012}.

Participation bias is not restricted to scenarios where the disease is life-threatening. For example, in electronic health record based studies, non-ignorable data incompleteness is a widespread concern \cite{haneuse2016}. In general, not only cases may be prone to different reasons not to participate in genotype data collection, but healthy controls may also be responsible for participation bias because they are less motivated to contribute genetic information to a study. More often, reasons for missingness are unknown. Therefore, an investigation on whether missingness is related to the genes of interest and an appropriate statistical method to adjust for potential participation bias are necessary. We face a related issue in the Two Sister Study \cite{o2016}, a family-based case-control genetic association study of breast cancer where genotype data for the single-nucleotide polymorphisms (SNPs) near the gene TOX3 on chromosome 16 were missing for $62\%$ of the sister-matched 924 controls, but at least one parent's genotype data were available for them. Standard statistical methods for dealing with missing data issues, such as multiple imputation \cite{allison2000,rubin1996} and maximum likelihood methods \cite{breslow1997}, are not applicable when missingness depends on the missing genotypes. 

Despite the widespread concern in genetic association studies, there has been a lack of methods to address participation bias in statistical inference. Modeling or adjusting for participation bias is challenging due to the dependence of missingness on the unobserved variables \cite{aschengrau2013,haneuse2016,haneuse2011}. To study a rare disease, when genotype data were incomplete only for cases, Chen and colleagues developed a family-supplemented design that allows adjustment for participation bias by collecting additional genotype data from family members of the cases \cite{chen2016}. In their design, family members, such as parents and children, provide information for a subject's missing genotypes following Mendelian genetic laws. A maximum likelihood approach was proposed to estimating the marginal effect of SNPs. In this work, we extend the family-supplemented design to general settings where controls may also have incomplete genotype data and the association model could include the main effects of SNPs, covariates, and gene-environment interaction effects. This extension, while allowing comprehensive association analysis, faces analytical challenges owing to possible correlations between genotype and covariates beyond the non-ignorable missingness. We refer to our method as the family-supplemented weighted empirical likelihood method (FS-WEL).

The rest of this paper is organized as follows. In section~\ref{methods}, we describe FS-WEL and provide the large sample theory. In section~\ref{simulation}, we evaluate the finite sample performance of FS-WEL via simulation studies in various scenarios. In section~\ref{realdata}, we demonstrate the proposed method by applying it to a young-onset breast cancer genetic association study. We make final remarks in Section~\ref{discussion}.


\section{Methods}
\label{methods}
\subsection{Framework and Notation}
Let $Y$ denote the binary phenotype variable with $Y = 1$ indicating a case and $Y = 0$ a control, $\vec{X}$ denote a vector of covariates, and $G$ denote the genotype of interest for a di-allelic SNP $A/a$ that is coded as the minor allele count, 0, 1, or 2. A logistic regression model is used to describe the relationship between $Y$, $\vec{X}$, and  $G$,
\begin{align*}
	\text{logit} \,\P(Y=1|\vec{X}, G)= \beta_0 + f_{\beta}(\vec{X}, G),
\end{align*}
where $f_{\beta}(\cdot)$ is a pre-specified function of $\vec{X}$ and $G$. We refer to this model as the association model. In this work, we consider a log-additive model for $G$ as $f_{\beta}(\vec{X}, G) = \vec{X}^T\vec{\beta}_1 + G\beta_2$, where $\vec{\beta}_1$ and $\beta_2$ are log odds ratio parameters of interest. Let $\theta$ denote the minor allele frequency. Under the Hardy-Weinberg Equilibrium \cite{good1978}, the frequency of $G$ is determined by $\theta$ as
\begin{align*}
	\P_{\theta}(G=g) = \theta^g(1-\theta)^{2-g}\cdot\{1+I(G=1)\},
\end{align*}
where $I(\cdot)$ is an indicator function. We make a rare disease assumption and thus the distribution of genotypes in controls can be approximated by that in the full population, i.e., $\P(G|Y=0) \approx \P_{\theta}(G)$. Define $\vec{\beta}=(\vec{\beta}^T_1,\beta_2)^{T}$ and $\eeta=(\vec{\beta}_1^T, \beta_2,\theta)^{T}$. Data for $(Y, \vec{X})$ are available for $N_1$ cases and $N_0$ controls. Let $N$ denote the total number of subjects, i.e., $N = N_0 + N_1$, and $\vec{x}^j$ $(j=1,...,J)$ denote the $j$th unique vector of values taken by $\vec{X}$. Genotype data $G$ are available for a subset of $n_1$ ($n_1 < N_1$) cases and $n_0$ ($n_0 < N_0$) controls. Let $R$ denote a subject's genotype data availability ($R=1$: available; $R=0$: not available). We refer to this subset of data with both $\vec{X}$ and $G$ for a particular SNP available (i.e., $R=1$) as complete observations, and $n=n_1+ n_0 = \sum_{i=1}^{N_1+N_0} R_i$ as the number of complete observations. We use a logistic regression model to describe the missingness mechanism as
\begin{align}
	\label{eq:missmodel}
	\text{logit}\,\P_{\alpha}(R=1|Y,\vec{X}, G)= h_{\alpha}(Y,\vec{X},G),
\end{align}
where $h_{\alpha}(\cdot)$ is a parametric function of $Y$, $\vec{X}$, and $G$. We refer to model~(\ref{eq:missmodel}) as the missingness model. Let $\vec{\alpha}$ denote the vector of all parameters in the missingness model. This model recognizes that missingness can depend on genotype itself. 

A non-ignorable missingness structure renders the imputation-based methods for estimating association parameters infeasible, and fitting the association model restricted to complete observations will result in biased estimates for parameters in vector $\bbeta$. Methods based on inverse probability weighting are standard approaches to consistent estimation using incomplete data. However, $G$ is not available for individuals whose outcome $R$ is 0, which makes it infeasible to estimate the probability of missingness by standard logistic regression analysis. Hence, we aim to develop a method that allows one to apply the weighting technique under non-ignorable missingness. 

Genotype data from family members can be used to provide information about an individual's missing genotype. This association motivated a family-supplemented design \cite{chen2016}, where genotype data from family members were collected as proxies for missing autosomal genotypes. Suppose that genotype data of a subject's spouse, denoted by $G^s$ and, of a child, denoted by $G^c$, can be made available for every subject whose genotype data are missing. Let $G^f$ denote the collection of all available familial genotype data $\{G^s,G^c\}$. We make the implicit assumption that genotypes have no influence on the availability of genetic information for family members. The key step is to compensate for the missingness of $G$ by incorporating $G^f$ into the estimating equation for the missingness model by inferring its distribution conditional on $G^f$, $\P_{\theta}(G|G^f)$, a defined function of the minor allele frequency. Let $G_i$ denote the $i$th subject's genotype, $G^f_i$ the vector of the $i$th subject's familial genotypes, and $\vec{X}_i$ the vector of the $i$th subject's covariates with $i=1,..., N$. Define $\delta_{\vec{x}g}$ as the point mass of $\P(\vec{X}=\vec{x}|G=g, Y=0)$.  Define $\ddelta=\{\delta_{\vec{x}g}; \vec{x}=\vec{x}^1,...,\vec{x}^{J-1} \text{ and } g=0, 1, 2\}$ and $\delta_{\vec{x}^Jg}=1-\sum_{j=1}^{J-1} \delta_{\vec{x}^{j}g}$. Below we present the proposed estimating equations and the inference procedure.

\subsection{Empirical Likelihood of the Association Model}
Had genotype data been obtained for all $N$ subjects in the case-control sample, the likelihood could have been explicitly written as a function of $\vec{\eta}$ and the vector of the nuisance conditional probabilities $\ddelta$, i.e.,
\begin{align*}
	L(\eeta, \ddelta)&=\prod\limits_{i=1}^{N_0} \P(\vec{X}_i,G_i|Y=0) \prod\limits_{i=N_0+1}^N \P(\vec{X}_i,G_i|Y=1). 
\end{align*}
Using a result from Satten and Kupper \cite{satten1993}, the joint distribution of $\vec{X}$ and $G$ for cases is related to that for controls as
$$
\P(\vec{X},G|Y=1)=\frac{e^{f_{\beta}(\vec{X},G)}\P(\vec{X}, G|Y=0)}{\sum\limits_{\vec{x}} \sum\limits_{g}e^{f_{\beta}(\vec{X},G)}\P(\vec{X}, G|Y=0)}.
$$
Then the above likelihood can be written as 
\begin{align*}
	\prod\limits_{i=1}^{N_0} \P(G_i|Y=0)\P(\vec{X}_i|G_i,Y=0)  \prod\limits_{i=N_0+1}^N \frac{e^{f_{\beta}(\vec{X}_i,G_i)}\P(G_i|Y=0)\P(\vec{X}_i|G_i,Y=0)}{\sum\limits_{\vec{x}} \sum\limits_{g}e^{f_{\beta}(\vec{X}, G)} \P(G|Y=0) \P(\vec{X}|G,Y=0)}.
\end{align*}
The intercept $\beta_0$ falls out of the likelihood function in this formulation. Therefore, we have avoided estimating probability $\P(Y|\vec{X},G)$, which is not estimable using a case-control sample. We then approximate $\P(G|Y=0)$, the distribution of genotype in controls, by $\P_{\theta}(G)$ under the rare disease assumption. We obtain the empirical likelihood by replacing $\P(\vec{X}|G, Y=0)$ with the point mass $\delta_{\vec{x}g}$,
\begin{align*}
	\prod\limits_{i=1}^{N} \P_{\theta}(G_i)\delta_{\vec{x}_i g_i}  \bigg\{ \frac{e^{f_{\beta}(\vec{X}_i,G_i)}}{\sum\limits_{\vec{x}} \sum\limits_{g}e^{f_{\beta}(\vec{X}, G)} \P_{\theta}(G) \delta_{\vec{x}g}}  \bigg\}^{Y_i}.
\end{align*}
Here, instead of treating the genotype frequency $\P(G|Y=0)$ as the probability of a multinomial variable, we model it as a parametric function of the minor allele frequency $\theta$ as governed by the population genetic theory, assuming the Hardy-Weinberg Equilibrium \cite{good1978}. This modeling, however, is not for the purpose of improving statistical efficiency. Instead, we model it to be consistent with a later step in our approach where the conditional probability $\P_{\theta}(G|G^f)$ needs to be estimated as a function of $\theta$ to infer the distribution of an individual's missing genotype. This point will be further discussed in section~\ref{missingness}, equation~(\ref{eq:alpha}).  

Maximizing the above empirical likelihood jointly with respect to all parameters, $\vec{\beta}$, $\theta$, and $\ddelta$, leads to a more efficient estimator of $\vec{\beta}$ than that from a standard logistic regression analysis due to the parametric modeling of $\P(G|Y=0)$. In standard logistic regression analysis where $P(G|Y=0)$ is treated as multinomial probabilities, the closed-form profile likelihood function for $\vec{\beta}$, obtained by maximizing the above empirical likelihood with respect to $\ddelta$ with $\vec{\beta}$ kept fixed, can greatly facilitate computation. We find that it is infeasible to derive such a closed-form profile-likelihood when $P(G|Y=0)$ is modeled parametrically in our study. Therefore, we consider the following procedure to reduce computational cost under the rare disease assumption. For each combination of $(\vec{x}, g)$, we estimate $\delta_{\vec{x}g}$ nonparametrically by the ratio of corresponding cell counts among controls, i.e.,
\begin{align}
	\label{eq:comdelta}
	\hat{\delta}_{\vec{x}g}=\frac{\sum\limits_{i=1}^{N_0} I(\vec{X}_i=\vec{x}, G_i=g)}{\sum\limits_{i=1}^{N_0} I(G_i=g)},
\end{align}
where $g=0,1,2$ and $\vec{x}=\vec{x}^1,...,\vec{x}^{J-1}$, and 
\begin{align*}
	\hat{\delta}_{(\vec{x}^J) g} = 1- \sum_{j=1}^{J-1} \hat{\delta}_{\vec{x}^{j}g}.
\end{align*}

\subsection{Estimating Equations for the Association Model}
Because the missingness is non-ignorable, in order to form an unbiased estimating equation for the association parameters, we use a weighting technique. More specifically, we first modify $\hat{\delta}_{\vec{x}g}$ in equation~(\ref{eq:comdelta}) by weighting complete controls using the inverse probability of genotype availability
\begin{align}
	\label{eq:delta}
	\hat{\delta}_{\vec{x}g}(\vec{\alpha})  &= \frac{\sum\limits_{i=1}^{N_0} \frac{I(R_i=1, \vec{X}_i=\vec{x}, G_i=g) }{ \P_{\alpha}(R=1|Y=0, \vec{X}=\vec{x}_i, G=g_i) } }
	{ \sum\limits_{i=1}^{N_0} \frac{I(R_i=1, G_i=g)}{\P_{\alpha}(R=1|Y=0, \vec{X}=\vec{x}_i, G=g_i)} },
\end{align}
and define $\hat{\ddelta}(\aalpha)=\{\hat{\delta}_{\vec{x}g}(\aalpha); \vec{x}=\vec{x}^1,...,\vec{x}^{J-1} \text{ and } g=0, 1, 2\}$. Then we weight a complete observation's empirical score function, $U_i\left(\eeta,\hat{\ddelta}(\aalpha)\right)$, with respect to parameter $\vec{\eta}=(\vec{\beta}_1^T,\beta_2,\theta)^T$ in the same way, that is
\begin{align}
	\label{eq:eta}
	\sum_{i=1}^{N} \frac{R_i}{\pi_i(\aalpha)} U_i\left (\eeta, \hat{\ddelta}(\aalpha) \right ) = \vec{0},  
\end{align}
where  $\pi_i(\aalpha)=\P_{\alpha}(R=1|Y_i,\vec{X}_i,G_i)$, 
\begin{align*}
	U_i\left( \eeta,\hat{\ddelta}(\aalpha) \right)
	= \begin{bmatrix}
		Y_i \left \{ \vec{X}_i-\frac{\sum_{\vec{x}}\sum_{g}Q \cdot \vec{X} }{\sum_{\vec{x}}\sum_{g} Q } \right \} \\
		Y_i \left \{ G_i-\frac{\sum_{\vec{x}}\sum_{g}  Q \cdot G }{\sum_{\vec{x}}\sum_{g} Q } \right \} \\
		\left( \frac{G_i}{\theta} - \frac{2-G_i}{1-\theta} \right) - Y_i \left\{  \frac{   \sum_{\vec{x}}\sum_g  Q (\frac{G}{\theta}-\frac{2-G}{1-\theta})   }{  \sum_{\vec{x}}\sum_g Q  }      \right\}  \\
	\end{bmatrix},
\end{align*}
and $Q=\hat{\delta}_{\vec{x}g}(\aalpha)\theta^G(1-\theta)^{2-G}\{1+I(G=1)\}e^{\vec{X}^T\vec{\beta_1}+G\beta_2}$.

\subsection{Estimating Equations for the Missingness Model}
\label{missingness}
Under non-ignorable missingness, it is challenging to estimate the vector $\aalpha$ in the missingness model, which is required for calculating weights in estimating equations~(\ref{eq:delta}) and (\ref{eq:eta}). Although the case-control sample can be treated prospectively with respect to estimation of $\aalpha$, since the outcome $Y$ is involved only as a predictor in the missingness model, one cannot directly fit a logistic regression model defined by equation~(\ref{eq:missmodel}) because $G$ is not available for individuals with $R=0$. Therefore, we propose an estimating equation that sums up the expectation of each observation's likelihood score function conditional on one's observed data $\vec{b}_i = (Y_i, \vec{X}_i, G^f_i, R=0)$ if $G_i$ is missing or $(Y_i, \vec{X}_i, G_i, R=1)$ if $G_i$ is available. Let $S_i(\vec{\alpha})$ be the standard score function for the logistic missingness model~(\ref{eq:missmodel}),
\begin{align*}
	S_i(\vec{\alpha})=\vec{d}_i \left \{R_i-\P_{\alpha}(R=1|Y_i,\vec{X}_i,G_i) \right \},
\end{align*}
where $\vec{d}_i$ is a vector of all covariates, and interaction terms in the missingness model \cite{agresti2011}. Our estimating equation is defined as follows:
\begin{align*}
	U^m \left (\vec{\alpha}, \eeta, \hat{\ddelta}(\aalpha) \right ) & = \sum_{i=1}^{N} E \left ( S_i(\vec{\alpha})|\vec{b}_i \right ) \\
	&=\sum_{i=1}^{N} \Big\{ R_iS_i(\aalpha)+(1-R_i)E\big(S_i(\aalpha)|\vec{b}_i \big)  \Big\}=\vec{0}.
\end{align*}
It is an unbiased estimating equation because its expectation reduces to the expectation of a regular score function and thus equals $0$. Moreover, by conditioning on the fully observed data, familial genotype data $G^f$ have been introduced into the estimation procedure. The estimation proceeds as follows. First, similar to the estimating equation for the association parameters, we relate the joint distribution of $(\vec{X}, G)$ among cases to that among controls in order to avoid estimating $\P(Y|\vec{X},G)$ under retrospective sampling \cite{satten1993}. Second, we model the genotype distribution as a parametric function of $\theta$ and thus estimation of $\P_{\theta}(G|G^f)$ is feasible. Finally, we explicitly write out the estimating equation for the missingness model as 
\begin{align}
	\label{eq:alpha}
	U^m \left (\aalpha, \eeta, \hat{\ddelta}(\aalpha) \right ) =&\sum_{i=1}^{N} \bigg\{R_i\vec{d}_i(1-\pi_i(\aalpha)) \\ \nonumber
	&- (1-R_i)
	\frac{\sum \limits_{g}  \vec{d}_i \P_{\alpha}(R=1|Y_i,\vec{X}_i,G) T_i}
	{ \sum \limits_{g} T_i} \bigg\} = 0,
\end{align}
where $T_i=\P_{\alpha}(R=0|Y_i,\vec{X}_i,G) e^{Y_i f_{\beta}(\vec{X}_i,G)}\hat{\delta}_{\vec{x}_ig}(\aalpha)\P_{\theta}(G_i^f)\P_{\theta}(G|G_i^f)$. This expression reformulates the challenging task of estimating $S_i(\aalpha)$ for an individual without genotype data into the key step of estimating $\P_{\theta}(G|G^f)$, which informs each missing genotype using familial genetic information. Consistent with the estimating equation for the association model, nuisance parameters in vector $\ddelta$ are estimated nonparametrically among complete controls by equation~(\ref{eq:delta}). Finally, taking expectation with respect to the missing variable $G$ integrates it out of the estimating equation. Detailed derivation under the rare disease assumption is shown in the supplementary material. Conditional on outcome $Y$, covariate $\vec{X}$, and genotype $G$, the probability of genotype availability does not depend on family member's genotype $G^f$, i.e., $\P(R |Y,\vec{X},G,G^f)=\P(R|Y,\vec{X},G)$. We refer to Supplementary Table 1 of Chen et al. \cite{chen2016} for the joint distribution of $(G, G^f)$.

\subsection{Asymptotic Properties}
We obtain point estimates ($\hat{\vec{\eta}}$, $\hat{\vec{\alpha}}$, $\hat{\vec{\delta}}(\hat{\vec{\alpha}})$) by jointly solving estimating equations~(\ref{eq:delta}), (\ref{eq:eta}), and (\ref{eq:alpha}). The expectations of these estimating functions, upon setting all parameters at the true values, equal zero, implying the consistency of ($\hat{\vec{\eta}}$, $\hat{\vec{\alpha}}$, $\hat{\vec{\delta}}(\hat{\vec{\alpha}})$) under regularity conditions \cite{van1998}. The asymptotic normality can then be established following the standard Z-estimation theory as follows. Define a constant matrix $$M = \begin{bmatrix}
	C_1 & C_2  \\ 
	D_1 & D_2 \\   
\end{bmatrix}, $$
where $C_1$, $C_2$, $D_1$, and $D_2$ are defined as
\begin{eqnarray*}
	C_1&=&E_{\{Y,\vec{X},G\}} \left( \frac{\partial U_i(\vec{\eta}, \ddelta )}{\partial \vec{\eta}} \right), \\
	C_2&=&E_{\{Y,\vec{X},G\}} \left( \frac{\partial U_i(\vec{\eta}, \ddelta )}{\partial \ddelta} \frac{\partial \hat{\ddelta}(\aalpha) }{\partial \vec{\alpha} } \right)-E_{\{Y,\vec{X},G\}} \left(U(\vec{\eta},\ddelta) \frac{\partial \text{log}\pi_i(\vec{\alpha})}{\partial \vec{\alpha} }  \right), \\
	D_1&=&E_{\{Y,\vec{X},G, G^f\}}\left(\frac{\partial U_i^m(\vec{\alpha},\ddelta,\vec{\eta}) }{\partial \vec{\eta}} \right), \\ D_2&=&E_{\{Y,\vec{X},G, G^f\}}\left(\frac{\partial U_i^m(\vec{\alpha},\ddelta,\vec{\eta}) }{\partial \vec{\alpha}} \right) +E_{\{Y,\vec{X},G,G^f\}}\left( \frac{\partial U_i^m(\vec{\aalpha},\ddelta,\vec{\eta}) }{\partial \vec{\delta}} \frac{ \partial \hat{\ddelta}(\aalpha)}{ \partial \aalpha }   \right ). 
\end{eqnarray*}
Expectations $\text{E}_{\{Y,\vec{X},G\}}(\cdot)$ and  $\text{E}_{\{Y,\vec{X},G,G^f\}}(\cdot)$  are taken with respect to joint distributions $\P(Y,\vec{X},G)$ and $\P(Y,\vec{X},G,G^f)$ in the case-control sample, respectively. These expected values can be estimated empirically using corresponding sample means. Then we define $\vec{a}_i=R_i U_i(\vec{\eta}, \ddelta)/\pi_i(\vec{\alpha})$ and $\vec{b}_i=\EEalpha$. For each combination of $(\vec{x}, g)$, we define 
\[
f^{\vec{x}g}_i = \frac{I(R_i=1, \vec{X}_i=\vec{x}, G_i=g)}{c_i} -\delta_{\vec{x}g},
\]
where $c_i=\P(R=1|Y=0,\vec{x}_i,g_i) \P(G=g|Y=0)$. The estimating equation for the nuisance parameter $\delta_{\vec{x}g}$ can then be expressed as 
\[
\hat{\delta}_{\vec{x} g}(\vec{\alpha})-\delta_{\vec{x}g}=\frac{1}{N_0} \sum_{i=1}^{N_0} f^{\vec{x}g}_i.
\]
Let $\vec{f}_i$ denote the vector of $\{f^{\vec{x}g}_i, \vec{x}=\vec{x}^1,\vec{x}^2,...,\vec{x}^{J-1} ~\text{and}~ g=0, 1, 2\}$. Define three matrices $V_{11}$, $V_{22}$, and $V_{12}$ as follows:
\begin{eqnarray*}
	V_{11} &=&\Cov(\vec{a}_i)+ p_0^{-1}\Cov_{0}(C_3 \vec{f}_i)+ \Cov_{0}(\vec{a}_i, C_3 \vec{f}_i) + \Cov_{0}(C_3 \vec{f}_i, \vec{a}_i),\\
	V_{22} &=& \Cov(\vec{b}_i)+p_0^{-1}\Cov_{0}(D_3 \vec{f}_i) + \Cov_{0}(\vec{b}_i, D_3 \vec{f}_i) + \Cov_{0}(D_3 \vec{f}_i,\vec{b}_i), \\
	V_{12} &=&\Cov(\vec{a}_i,\vec{b}_i)+\Cov_{0}(C_3 \vec{f}_i,\vec{b}_i ) + \Cov_{0}(\vec{a}_i,D_3 \vec{f}_i) + p_0^{-1} \Cov_{0}(C_3\vec{f}_i, D_3\vec{f}_i) ,
\end{eqnarray*}
where constants $C_3$ and $D_3$ are defined as
\begin{eqnarray*}
	C_3&=&E_{\{Y,\vec{X},G\}} \left(  \frac{\partial U_i(\vec{\eta}, \ddelta ) }{\partial \ddelta} \right),  \\
	D_3&=&E_{\{Y,\vec{X},G,G^f\}}\left(\frac{\partial U_i^m(\vec{\alpha},\ddelta,\vec{\eta}) }{\partial \ddelta} \right).
\end{eqnarray*}
Covariance $\Cov_{0}$ is taken with respect to the joint probability of $(\vec{X}, G)$ within controls, i.e., $\P(\vec{X}, G|Y=0)$, and $\Cov$ is taken with respect to the joint probability $\P(Y,X,G)$ in the full case-control sample. Covariance-variance matrices $\Cov_{0}$ and $\Cov$ can be consistently estimated empirically using corresponding covariances within controls and in the full sample, respectively. Finally we define a matrix 
$$V = \begin{bmatrix}
	V_{11} & V_{12}  \\
	V_{12}^T & V_{22}  \\
\end{bmatrix}.$$ 
The asymptotic normality of ($\widehat{\vec{\eta}}^T, \widehat{\vec{\alpha}}^T)^T$ is summarized in the following corollary, with detailed proof provided in the supplementary material.

\begin{Corollary}
	Suppose that the true values of $\eeta$ and $\aalpha$ lie inside a compact space, and that components of $(\vec{X}, G )$ are bounded. Under regularity conditions \cite{van1998}, $(\hat{\eeta}^T, \hat{\aalpha}^T)^T$ is asymptotically normally distributed with  
	\begin{align*}
		\sqrt N \begin{bmatrix}
			\hat{\eeta}-\eeta \\
			\hat{\aalpha} -\aalpha \\
		\end{bmatrix} \xrightarrow{D}
		N \left(\begin{bmatrix} 
			\vec{0} \\
			\vec{0} \\
		\end{bmatrix},  M^{-1}V(M^{-1})^T  \right).
	\end{align*} 
\end{Corollary}

\subsection{Approach Implementation}
Since analytical solutions to estimating equations~(\ref{eq:delta}), (\ref{eq:eta}), and (\ref{eq:alpha}) do not exist, we apply an iterative method to find numerical solutions as below.
\begin{enumerate}
	\item Set initial values of $\vec{\eta}$ and $\vec{\alpha}$, i.e., $\hat{\vec{\eta}}^0$ and $\hat{\vec{\alpha}}^0$. 
	\item Calculate the initial value of $\hat{\delta}_{{\vec{x}}{g}}(\vec{\alpha})$ in equation~(\ref{eq:delta}) using $\hat{\vec{\alpha}}^0$, i.e., $\hat{\delta}_{{\vec{x}}{g}}(\hat{\vec{\alpha}}^0)$.
	\item Keep $\hat{\vec{\eta}}^0$ fixed in equation~(\ref{eq:alpha}), and solve for $\vec{\alpha}$ using the Newton's method to obtain an updated estimate $\hat{\vec{\alpha}}^1$. 
	\item Update $\hat{\delta}_{{\vec{x}}{g}}(\vec{\alpha})$ in equation~(\ref{eq:delta}) using $\hat{\vec{\alpha}}^1$, i.e., $\hat{\delta}_{{\vec{x}}{g}}(\hat{\vec{\alpha}}^1)$.
	\item Keep $\hat{\vec{\alpha}}^1$ and thus $\hat{\delta}_{{\vec{x}}{g}}(\hat{\vec{\alpha}}^1)$ fixed in equation~(\ref{eq:eta}) and solve for $\vec{\eta}$ to update its value, i.e., $\hat{\vec{\eta}}^1$.
	\item Repeat step 3, 4, and 5 until consistency is achieved.
\end{enumerate}
There is no best way to set initial values at step 1. In simulation studies,  we used naive estimates of $\vec{\eta}$ based on complete observations, assuming missing completely at random, as initial values. More details are described in section~\ref{simulation}.  We set $\hat{\vec{\alpha}}^0$ to $\vec{0}$ in simulations. R package ``nleqslv" can be used to solve the system of nonlinear equations at step 3 and 5.

\section{Results}
\subsection{Simulation Study}
\label{simulation}
We considered different outcome prevalences, minor allele frequencies, proportions of missingness, and effect sizes of genotype in the association and missingness models.  We included genotype $G$ and a binary covariate $X$ in the association model. We first generated $G$ for each individual in a cohort of size $10^6$ based on minor allele frequency $\theta=0.2$ and $\theta=0.5$ under the assumption of the Hardy-Weinberg Equilibrium\cite{good1978}. That is, $G$ was generated to take on values 0, 1, and 2 from a multinomial distribution with frequencies $(1-\theta)^2$, $2\theta(1-\theta)$, and $\theta^2$, respectively. Then we generated genotype $G^s$ for every individual's spouse in the same way as for $G$ and genotype $G^c$ for each couple's child from a multinomial distribution of probabilities derived under the same assumptions (Supplementary Table 1). Then we generated a binary variable $X$ with levels 0 and 1 conditional on $G$ from a Bernoulli distribution of probability $\P(X=1|G=0)=0.3$, $\P(X=1|G=1)=0.5$, and $\P(X=1|G=2)=0.35$.  The binary outcome variable $Y$ was then generated based on the logistic regression model
\begin{align}
	\label{eq:main}
	\text{logit}\,\P(Y=1|X, G) = \beta_0+\beta_1X+\beta_2G,
\end{align}
where $\beta_1=\log(1.2)$ and $\beta_2$ was set to $\log(1.2)$ and $\log(1.5)$ corresponding to relatively weak and strong associations between the outcome and genotype. Intercept $\beta_0$ was chosen to achieve outcome prevalence of $3\%$ and $10\%$.

We then generated $R$, the missingness status of genotype, from model
\begin{align*}
	\text{logit}\,\P_{\alpha}(R=1|Y, X, G)=\alpha_0 + \alpha_1 Y + \alpha_2 X + \alpha_3 G +  \alpha_4 YX+ \alpha_5 YG,
\end{align*}
where $\alpha_1=\log(0.6)$, $\alpha_2=\log(1.2)$, and $\alpha_3$ took on $\log(1.2)$ and $\log(1.5)$ to simulate relatively weak and strong associations between missingness and genotype. We considered non-differential (i.e., $\alpha_4=\alpha_5 =0$) and moderately differential (i.e., $\alpha_4 = \alpha_5 = \log(1.5)$) associations of missingness and $X$, and of missingness and $G$ between cases and controls. Hereafter, we will refer to these two models as the non-differential and differential missingness models. Intercept $\alpha_0$ was selected to determine $80\%$ and $60\%$ genotype availabilities. We randomly selected 2000 cases and 2000 controls from the cohort to form a case-control sample.

Naive estimates of $\beta_1$ and $\beta_2$ were obtained under two missing mechanisms. Assuming missing completely at random (MCAR), we obtained these estimates by fitting logistic regression model~(\ref{eq:main}) directly to complete observations. The naive estimate of $\theta$, denoted as $\tilde \theta$, was calculated as
\begin{align}
	\label{eq:maf}
	\tilde \theta = \frac{2n_{02}+n_{01}}{2n_0},
\end{align}
where $n_{01}$ and $n_{02}$ were numbers of individuals who had one and two copies of the minor allele, respectively, among $n_0$ controls whose genotype data were available. When missing at random (MAR) was wrongly assumed where missingness did not depend on genotype $G$, we fit the following missingness model
\begin{align*}
	\text{logit}\,\P_{\alpha}(R=1|Y, X)=\alpha_0 + \alpha_1Y + \alpha_2X
\end{align*}
on the full case-control sample to estimate the probability of genotype availability for each observation. We then weighted each complete observation by the reciprocal of this probability in logistic regression model~(\ref{eq:main}) to estimate $\beta_1$ and $\beta_2$. We estimated $\theta$ among complete controls as
\begin{align*}
	\tilde \theta=\frac{\sum_{i=1}^{N_0} \frac{I(R_i=1,G_i=1)+2I(R_i=1,G_i=2)}{P(R=1|Y=0,X=x_i)} }{\sum_{i=1}^{N_0} \frac{2I(R_i=1)}{P(R=1|Y=0, X=x_i)}  }.
\end{align*}
For each set of the parameters, we repeated the above steps 1000 times. We did not consider the original FSD method \cite{chen2016} in simulation study because it cannot handle missingness of genotype data among both cases and controls or adjust for the effect of covariate $X$ in the association model~(\ref{eq:main}). 

Table~\ref{tab:est} presents the estimates of the log odds ratio parameter $\bbeta$ in the association model and minor allele frequency using FS-WEL under outcome prevalence of $3\%$ and minor allele frequency of 0.2. Point estimates are all close to true parameter values, the empirical and asymptotic standard errors are close, and the coverage probabilities are consistent with the nominal level of 0.95 across all simulated scenarios. Similar results for parameters in the missingness model are presented in Supplementary Table 2.

\begin{table}[ht]
	\small
	\centering
	\caption{The estimated log odds ratio with respect to covariate $X$ ($\hat \beta_1$) and genotype $G$ ($\hat \beta_2$) in the association model and the estimated minor allele frequency ($\hat \theta$) using the weighted empirical likelihood method. The true values of $(\alpha_3,\alpha_4,\alpha_5)$ in the three missingness models are: weak $=(0.182, 0.405, 0.405)$, strong $=(0.405, 0.405, 0.405)$, and non-differential $=(0.405, 0, 0)$. The true value of $\beta_2$ and all estimates are presented in the magnitude of $10^{-3}$.} 
	\label{tab:est}
	\begin{tabular}{ccccccccc}
		\toprule
		$P(R=1)$ & $\beta_2$ & missing & \multicolumn{2}{c}{$\hat{\beta}_1$}  & \multicolumn{2}{c}{$\hat{\beta}_2$}  & \multicolumn{2}{c}{$\hat{\theta}$}\\ 
		\cmidrule(lr){4-5} \cmidrule(lr){6-7} \cmidrule(lr){8-9}
		&								&				 & est(asy/emp)  & cov  & est(asy/emp) & cov & est(asy/emp) & cov   \\ \midrule
		0.8 & 182  & weak & 180 (71/69)& 961 & 179 (59/62) & 928  & 200 (7/7)& 946  \\ 
		&         & strong & 180 (71/68)& 964 & 179 (58/57) & 957  & 200 (7/7)& 952  \\ 
		&         & ND & 185 (71/68)& 960 & 182 (60/61) & 947  & 199 (7/7)& 951  \\  \cmidrule(lrrrr){2-9}
		& 405  & weak & 183 (73/70)& 957  & 404 (57/58)& 945  & 198 (7/7)& 947   \\ 
		&         & strong & 180 (73/69)& 972 & 406 (56/60)& 934 & 198 (7/7)& 946  \\ 
		&         & ND & 181 (73/67)& 969  & 403 (58/62)& 935 & 198 (7/7)& 942   \\  \midrule
		0.6 & 182  & weak & 179 (75/68)& 967  & 179 (65/69) & 936 & 200 (8/8)& 950  \\ 
		& 	      & strong & 180 (75/68)& 969  & 179 (64/63) & 961 & 200 (8/8)& 953 \\ 
		&         & ND & 185 (75/68)& 963  & 175 (67/71)  & 937  & 200 (8/8)& 952 \\   \cmidrule(lrrrr){2-9}
		& 405  & weak & 183 (78/70)& 967   & 406 (63/64)& 946 & 198 (8/8)& 937 \\ 
		& 	      & strong & 179 (77/70)& 975   & 406 (62/66)& 927 & 198 (8/8)& 949  \\ 
		& 	      & ND & 182 (77/68)& 976  & 403 (64/69)& 934  & 198 (8/8)& 942   \\ 
		\bottomrule
	\end{tabular}
	\begin{tablenotes}
		\item{est: estimate; asy: asymptotic standard error; emp: empirical standard error; ND: non-differential, cov: coverage.}
	\end{tablenotes}
\end{table}

Table~\ref{tab:bias} presents the biases in the log odds ratio estimates in the association model and their mean squared errors (MSEs) in the same scenarios as in Table~\ref{tab:est}. Biases were calculated as the difference between the mean point estimates based on 1000 iterations and the true values of parameters. Biases are within $2\%$ of true parameter values in log odds ratio estimate $\hat{\beta_1}$, $4\%$ in $\hat{\beta_2}$, and $1\%$ in $\hat{\theta}$ in use of FS-WEL. In contrast, the naive method assuming MCAR overestimates $\beta_1$ by $47\%-95\%$ and $\beta_2$ by $18\%-82\% $ in scenarios of differential missingness models and by $8\%-15\%$ and $8\%-24\%$, respectively, when the missingness model is non-differential. It overestimates $\theta$ by $2\%-12\%$. Moreover, MSEs based on the naive method assuming MCAR are $1.3-8.1$ times greater for $\hat{\beta}_1$, $1.3-6.4$ times greater for $\hat{\beta}_2$, and $1.1-10.4$ times greater for $\hat{\theta}$ than those in the weighted empirical likelihood method. The results using the naive method assuming MAR are very similar to those under MCAR. Our proposed method is able to correct for non-ignorable missingness and yields unbiased estimates for log odds ratio and minor allele frequency parameters across all simulation settings even under severe missingness. Naively ignoring the non-ignorable missingness structure, however, may lead to large biases in these estimates. Biases and MSEs become even more prominent when the associations between missingness and covariates ($X$ and $G$) further differentiate between cases and controls and when the proportion of missing data increases. 

\begin{table}[ht!]
	\centering
	\small
	\caption{The bias and the mean squared error of estimated log odds ratio of covariate $X$ ($\hat \beta_1$) and genotype $G$ ($\hat \beta_2$) in the association model and the estimated minor allele frequency ($\theta$). The true values of $(\alpha_3,\alpha_4,\alpha_5)$ in the three missingness models are: weak $=(0.182, 0.405, 0.405)$, strong $=(0.405, 0.405, 0.405)$, and non-differential $=(0.405, 0, 0)$. In each of the 12 settings, true values and estimates of coefficients $\beta_1$, $\beta_2$ and $\theta$ are presented in this order in the magnitude of $10^{-3}$. }
	\label{tab:bias}
	\begin{threeparttable}
		\begin{tabular}{ccccrrrrrr}
			\toprule
			\multicolumn{3}{c}{Model}    &     & \multicolumn{2}{c}{MCAR}  &  \multicolumn{2}{c}{MAR} &  \multicolumn{2}{c}{FS-WEL} \\ 
			\cmidrule(lr){1-3} \cmidrule(lr){5-6} \cmidrule(lr){7-8} \cmidrule(lr){9-10}
			$\P(R=1)$  & $\beta_2$ &  Missing &True  & Bias & MSE &  Bias & MSE &  Bias & MSE\\ \midrule
			0.8 	& 182 & weak   & $\beta_1\ (182)$ & 91.0 & 14.3 & 85.9 & 13.4 & -2.7 & 4.8 \\ 
			&       &      & $\beta_2\ (182)$ & 71.7 & 9.0 & 72.4 & 9.1&  -3.5 & 3.9 \\ 
			&       &      & $\theta\phantom{_1}\  (200)$ & 4.5	& 0.1 & 5.9 & 0.1 & -0.5 & 0.1 \\ 
			&   & strong   & $\beta_1\  (182)$ & 86.4 & 13.2 & 82.1 & 12.5 & -2.6 & 4.6 \\ 
			&       &      & $\beta_2\  (182)$ & 69.1 & 8.4 & 70.1 & 8.6 & -3.4 & 3.3 \\ 
			&       &      & $\theta\phantom{_1}\  (200)$ & 11.1 & 0.2 & 12.3 & 0.2 & -0.3 & 0.1 \\ 
			&        & ND  & $\beta_1\  (182)$ & 18.5 & 6.5 & 8.5 & 6.1 & 3.1 & 4.6 \\ 
			&       &      & $\beta_2\ (182)$ & 36.3 & 4.9 & 36.3 & 4.9 & $<$0.1 & 3.8 \\ 
			&       &      & $\theta\phantom{_1}\ (200)$ & 10.1 & 0.2 & 10.6 & 0.2 & -0.9 & 0.1 \\  \cmidrule(lrrrrrr){2-10}
			& 405 & weak   & $\beta_1\ (182)$ & 91.5 & 14.5 & 89.0 & 14.0 & 0.4 & 4.9 \\ 
			&       &      & $\beta_2\ (405)$ & 75.3 & 9.1 & 76.1 & 9.2 & -1.0 & 3.3 \\ 
			&       &      & $\theta\phantom{_1}\ (200)$ & 3.0 & 0.1 & 4.3 & 0.1 & -1.9 & 0.1 \\ 
			&   & strong   & $\beta_1\ (182)$ & 88.8 & 14.1 & 87.1 & 13.1 & -2.5 & 4.9 \\ 
			&       &      & $\beta_2\ (405)$ & 73.9 & 9.1 & 74.8 & 9.2 & 0.8 & 3.6 \\ 
			&       &      & $\theta\phantom{_1}\ (200)$ & 8.9 & 0.1 & 10.1 & 0.2 & -2.1 & 0.1 \\ 
			&       & ND   & $\beta_1\ (182)$ & 14.4 & 5.9 & 5.5 & 5.7 & -0.9 & 4.5 \\ 
			&       &      & $\beta_2\ (405)$ & 34.1 & 4.9 & 34.2 & 5.0 & -2.4 & 3.8 \\ 
			&       &      & $\theta\phantom{_1}\ (200)$ & 9.1 & 0.1 & 9.5 & 0.1 & -1.9 & 0.1 \\  \midrule
			0.6 	& 182 & weak   & $\beta_1\ (182)$ & 169.6 & 37.1 & 165.0 & 35.5 & -3.8 & 4.6 \\ 
			&       &      & $\beta_2\ (182)$ & 149.6 & 27.1 & 152.0 & 27.8 & -2.9 & 4.7 \\ 
			&       &      & $\theta\phantom{_1}\ (200)$ & 9.2 & 0.2 & 11.9 & 0.2 & -0.5 & 0.1 \\ 
			&   & strong   & $\beta_1\ (182)$ & 161.0 & 33.9 & 157.8 & 32.9 & -2.5 & 4.6 \\ 
			&       &      & $\beta_2\ (182)$ & 137.3 & 23.6 & 139.8 & 24.2 & -3.2 & 4.0 \\ 
			&       &      & $\theta\phantom{_1}\ (200)$ & 23.2 & 0.6 & 25.5 & 0.7 & -0.3 & 0.1 \\ 
			&       & ND   & $\beta_1\ (182)$ & 27.2 & 9.3 & 14.5 & 8.7 & 3.0 & 4.7 \\ 
			&       &      & $\beta_2\ (182)$ & 44.4 & 7.3 & 44.5 & 7.3 & -7.2 & 5.1 \\ 
			&       &      & $\theta\phantom{_1}\ (200)$ & 22.8 & 0.6 & 23.7 & 0.6 & -0.3 & 0.1 \\  \cmidrule(lrrrrrr){2-10}
			& 405 & weak   & $\beta_1\ (182)$ & 172.2 & 37.9 & 170.8 & 37.4 & 0.5 & 4.9 \\ 
			&       &      & $\beta_2\ (405)$ & 147.8 & 26.3 & 149.9 & 27.0 & 0.5 & 4.1 \\ 
			&       &      & $\theta\phantom{_1}\ (200)$ & 8.0 & 0.1 & 10.6 & 0.2 & -2.0 & 0.1 \\ 
			&   & strong   & $\beta_1\ (182)$ & 160.9 & 33.8 & 161.1 & 33.8 & -3.0 & 4.9 \\ 
			&       &      & $\beta_2\ (405)$ & 139.8 & 24.3 & 141.9 & 24.9 & 0.7 & 4.4 \\ 
			&       &      & $\theta\phantom{_1}\ (200)$ & 21.0 & 0.5 & 23.3 & 0.6 & -2.0 & 0.1 \\ 
			&       & ND   & $\beta_1\ (182)$ & 19.4 & 8.9 & 7.3 & 8.5 & -0.6 & 4.6 \\ 
			&       &      & $\beta_2\ (405)$ & 46.6 & 7.2 & 46.7 & 7.3 & -2.5 & 4.7 \\ 
			&       &      & $\theta\phantom{_1}\ (200)$ & 21.0 & 0.5 & 21.9 & 0.6 & -1.8 & 0.1 \\ 
			\bottomrule
		\end{tabular}
		\begin{tablenotes}
			\item{FS-WEL: family-supplemented weighted empirical likelihood method; MCAR: missing completely at random; MAR: missing at random; MSE: mean squared error; ND: non-differential.}
		\end{tablenotes}
	\end{threeparttable}
\end{table}

We further investigated the performance of FS-WEL in scenarios with a higher minor allele frequency ($0.5$) and a greater prevalence ($0.1$). Our method still presents a superior capacity of correcting for biases in log odds ratio estimates compared with the naive method (Supplementary tables $3$-$8$). The good performance of our method under prevalence of 0.1 demonstrates its robustness with respect to the rare disease assumption, which renders this method useful in a broader range of studies. We also evaluated the performance of our proposed method when the assumption of MCAR held, and the simulations suggest that it produces consistent log odds ratio estimates and slightly improves the efficiency of the estimates in comparison with the naive method assuming MCAR (Supplementary tables $9$-$10$).

When the empirical distributions of covariates estimated only based on controls need to be applied to cases, extrapolation may be necessary because observed covariate values in cases and controls may differ, particularly when continuous covariates are involved. We implemented a simple smoothing technique to overcome this numerical issue. For each distinct covariate value denoted by $\vec{x}^*$ that is only observed in cases, we identified a value $\vec{x}^{**}$ in controls that is adjacent to $\vec{x}^*$. We then used $\delta_{\vec{x}^{**}g}/2$ as the estimated empirical probability at both $\vec{x}^*$ and $\vec{x}^{**}$, which ensured that the estimated empirical probability summed over all covariate values, $\sum_{\vec{x}} \delta_{\vec{x}g}$, equals one. We found in numerical studies that FS-WEL still successfully corrected for participation bias under this implementation. The issue of differential empirical domains between cases and controls is usually not worrisome when all covariates are discrete as in our real data example, particularly when the sample size is not too small.

\subsection{Real Data Example}
\label{realdata}
In this section, we analyzed a dataset derived from the Two Sister Study (\url{https://sisterstudy.niehs.nih.gov/English/twosisterstudy.htm}), a family-based study of young-onset (age under 50 years) breast cancer. In the original study, each family contributed a sister diagnosed with young-onset breast cancer, a control sister, and both their parents when possible. The original analysis \cite{o2016} estimated relative risks by applying a log-linear model \cite{weinberg1998} for case-parent data. 

Our method was proposed in a different setting, where a classical case-control design was used. We were interested in investigating whether genes that were known to be associated with the risk of young-onset breast cancer had different effects depending on other risk factors. To perform the case-control analysis of interest, we drew data from the Two Sister Study to construct a case-control sample where every individual had a first-degree family history of breast cancer. Most controls were not genotyped in the original family-based design, but because their parents' genotypes were available, these controls became informative using our proposed method. 

The case-control sample included 521 women who had been diagnosed with breast cancer before age 50 and had one or more first-degree family members previously diagnosed with breast cancer (often the mother), and 924 controls who had never been diagnosed with breast cancer, but had an affected sister. When multiple controls were available from the same family, we randomly selected one of them. Thus study subjects in this analysis by selection were unrelated, but some with missing genotypes had first-degree relatives available to provide proxy genotype information. Let $Y$ denote the breast cancer case-control status. 

Age at first full-term pregnancy ($X_1$) and age at menarche ($X_2$) are two standard risk predictors for breast cancer \cite{chen2006,gail1989}. For $X_1$, we use NL to denote nulliparous. The distributions of these two variables are provided in Supplementary Table 11. Genetic data for SNPs rs8050542, rs8046979, rs4784220, rs1420533 and rs43143 near gene TOX3 on chromosome 16 were available for all cases and 351 controls (38.0\%). Among the remaining 573 controls, genotype data were available for both parents of 328 controls and for one parent of 245 controls. We considered the following association model,
\begin{align*}
	\text{logit}\,\P(Y=1 | X_1, X_2, G) = &\beta_0+\beta_{11} I(24<X_1\leq30) +\beta_{12} I(X_1>30) + \beta_{13} I(\text{NL}) \\ \nonumber
	+&\beta_{21} I(12 < X_2 <14) +\beta_{22} I(X_2 \leq 12) +\beta_{3} G  \\ \nonumber
	+&\beta_{41} I(24<X_1\leq30)G +\beta_{42} I(X_1>30)G + \beta_{43} I(\text{NL})G,
\end{align*} 
to investigate whether age at first full-term pregnancy modifies the effect of $G$, the number of copies of the minor allele. It was uncertain whether the availability of $G$ depended on $G$ itself or age at first full-term pregnancy. Therefore, we considered a missingness model that flexibly allows non-ignorable missingness among controls. That is,
\begin{align*}
	\text{logit} \,\P(R=1 | Y=0, X_1, X_2, G) = &\alpha_0 + \alpha_{11} I(24<X_1\leq30) +\alpha_{12} I(X_1>30) \\
	+&\alpha_{13} I(\text{NL}) +\alpha_{21} I(12 < X_2 <14) \\
	+&\alpha_{22} I(X_2 \leq 12)  + \alpha_3 G.
\end{align*}
For each of the five SNPs, we applied our new method and investigated whether missingness depended on genotype and age at first full-term pregnancy at that locus and used parents' genotype data to help infer the distribution of a woman's missing genotype. In contrast, naive estimates were obtained from fitting the association model to the complete observations, and the minor allele frequency $\theta$ was estimated using complete controls based on equation~\ref{eq:maf} with standard error $(\tilde \theta (1-\tilde{\theta})/2n_0)^{1/2}$. Results of SNPs rs4784220 and rs1420533 are presented in Table~\ref{tab:twosis}. Similar results of SNPs rs8051542, rs8046979, and rs43143 and presented in Supplementary Table 12.

\begin{table}[ht]
	\small
	\centering
	\caption{Estimated log odds ratio parameters and the asymptotic standard errors in the association model of the Two Sister Study.}
	\label{tab:twosis}
	\begin{threeparttable}
		\begin{tabular}{lrcrccc}
			\toprule
			&  \multicolumn{4}{c}{Association Model}  & Missingness Model \\ \cmidrule(lrrr){2-5}  \cmidrule(lr){6-6}
			Model   & FS-WEL (SE) & \textit{p}-value  & Naive (SE)  & \textit{p}-value & \textit{p}-value \\ 
			\midrule
			rs4784220                            &0.369 (0.244)& 0.065 & 0.397 (0.214)& 0.032 & 0.462 \\ 
			$X_1$ (24, 30]                       &0.245 (0.325)& 0.226 & 0.112 (0.338)& 0.370 & 0.061 \\ 
			\phantom{$X_1$ }$>30$                &0.683 (0.364)& 0.030 & 0.710 (0.372)& 0.028 & 0.024 \\ 
			\phantom{$X_1$ }NL                   &0.802 (0.644)& 0.106 & 0.720 (0.333)& 0.015 & 0.182 \\ 
			$X_2$ (12, 14)                       &0.240 (0.208)& 0.125 & 0.367 (0.187)& 0.025 & 0.104 \\ 
			\phantom{$X_2$ }$\leq 12$            &0.363 (0.230)& 0.057 & 0.454 (0.176)& 0.005 & 0.119 \\ 
			Interaction (24, 30]                                  &-0.056 (0.290)& 0.423 & 0.048 (0.302)& 0.437 &  \\ 
			\phantom{Interaction }$>30$& -0.020 (0.353)& 0.478 &-0.037 (0.348)& 0.458 &  \\ 
			\phantom{Interaction }NL&-0.549 (0.553)& 0.160 &-0.542 (0.300)& 0.036 &  \\ 
			MAF &0.439 (0.008)&  & 0.439 (0.026)&  &  \\ \midrule
			rs1420533                            &-0.342 (0.221)& 0.061 &-0.351 (0.190)& 0.032 & 0.490 \\ 
			$X_1$ (24, 30]                       &0.190 (0.331)& 0.283 & 0.178 (0.330)& 0.295 & 0.060 \\ 
			\phantom{$X_1$ }$>30$                &0.722 (0.420)& 0.043 & 0.681 (0.388)& 0.039 & 0.024 \\ 
			\phantom{$X_1$ }NL                   &-0.421 (0.461)& 0.180 &-0.411 (0.312)& 0.094 & 0.189 \\ 
			$X_2$ (12, 14)                       &0.361 (0.210)& 0.043 & 0.392 (0.187)& 0.018 & 0.102 \\ 
			\phantom{$X_2$ }$\leq 12$            &0.382 (0.238)& 0.054 & 0.462 (0.176)& 0.004 & 0.119 \\ 
			Interaction (24, 30]                                  &-0.043 (0.272)& 0.437 &-0.008 (0.269)& 0.487 &  \\ 
			\phantom{Interaction }$>30$ & -0.127 (0.349)& 0.358 &-0.009 (0.305)& 0.488 &  \\ 
			\phantom{Interaction }NL &0.708 (0.501)& 0.079 & 0.692 (0.262)& 0.004 &  \\ 
			MAF &0.496 (0.011)&  & 0.489 (0.027)&  &  \\ 
			\bottomrule
		\end{tabular}
		\begin{tablenotes}
			\footnotesize
			\item{ FS-WEL: family-supplemented weighted empirical likelihood method; SE:  standard error; NL: nulliparous.}
		\end{tablenotes}
	\end{threeparttable}
\end{table}

Difference in the estimates between these two methods ranges from $1$\%-$54$\% for $\beta_{11}$, $1$\%-$9$\% for $\beta_{12}$, $2$\%-$15$\% for $\beta_{13}$, $9$\%-$54$\% for $\beta_{21}$, $18$\%-$30$\% for $\beta_{22}$ and $1$\%-$45$\% for $\beta_3$. Estimates of the parameters for interactions differ by $16\%$-$80\%$ for $\beta_{41}$, $22\%$-$93\%$ for $\beta_{42}$, and $1\%$-$47\%$ for $\beta_{43}$. The estimates of $\theta$ are very close ($1$\%-$3$\%). 

Missingness of genotype data in controls does not appear to depend on the SNPs under study ($p$-value is $0.36$-$0.49$), but is related to the covariate age at first full-term pregnancy ($p$-value $<0.05$, Table~\ref{tab:twosis}). Thus, missingness of the interactions between genotype and age at first full-term pregnancy is non-ignorable, which results in biases in the estimated log odds ratio for age at first full-term pregnancy and the interactions in the association model using the naive method. Biases in log odds ratio estimates for other variables might be caused by the correlation between age at first full-term pregnancy and other variables. 

Two of the five SNPs under study, rs4784220 ($p$-value $=0.065$) and rs1420533 ($p$-value $=0.061$) are marginally associated with the risk of young-onset breast cancer (Table~\ref{tab:twosis}); SNPs rs8051542 ($p$-value $=0.139$), rs8046979 ($p$-value $=0.106$), and rs43143 ($p$-value $=0.42$) are not statistically significantly associated (Supplementary Table 12). The associations do not seem to vary with age at first full-term pregnancy  ($p$-value: 0.079-0.478). The odds of developing young-onset breast cancer for women with first-degree breast cancer family history who gave birth to their first child after age 30 are $90$\%-$122$\% greater than those whose first child was born before age 24 ($p$-value: 0.028--0.043). The odds for those whose age at menarche was before 12 and between 12-14 tend to be $40$\%-$47$\% greater ($p$-value: 0.051-0.069) and $27$\%-$44$\% greater ($p$-value: 0.025-0.13) than those who had menarche after age 14, respectively.  Overall the results are consistent with those reported from the original Two Sister Study \cite{o2016,shi2017}.

\section{Discussion}
\label{discussion}
Participation bias can induce a non-ignorable missingness structure that is difficult to investigate or adjust for. With a family member serving as an informative proxy for the missing genotype data, our proposed method can effectively correct for participation bias in general scenarios where both cases and controls may have genetic information missing for reasons related to genotypes, covariates, and their interactive effects. In practice, genotype data may be missing for only cases or controls. Survival bias for studies of life-threatening diseases is one example. Our method can be applied with straightforward modification on the missingness model. When only cases are subject to missingness, the estimation can be further simplified because the nuisance parameter, the empirical distribution of the covariates conditional on the genotype, can be directly estimated using all controls (equation~\ref{eq:comdelta}). In general, the reasons for missingness may be unknown. Our method provides an approach to sensitivity analyses regarding whether a missing-at-random or missing-completely-at-random assumption holds. Furthermore, when the missingness is completely at random, our method showed improved statistical efficiency compared with the naive method, possibly due to parametric modeling of the genotype distribution and larger effective sample sizes. Our method is practical because it can flexibly utilize different types of familial genetic information, such as that from one parent, from children only, and possibly other relatives. Data from first-degree relatives are preferred because the use of strong proxies leads to more precise parameter estimates. 

Our work focused on scenarios where genotype data are missing due to incomplete participation. When covariate-driven participation bias is of concern, a proactive approach would be to collect additional variables that are possibly related to participation through a multiphase design \cite{haneuse2016,haneuse2011}. These auxiliary data would allow us to explore the missingness mechanism and consequently to apply appropriate methods for bias correction. Our method can be extended to address general missing data problems with non-ignorable missingness for genetic information and random missingness for other covariates. Incorporation of familial genetic data through similar weighting ideas based on available methods for random missingness of covariates might be feasible, which we hope to investigate in future work.

\section*{Acknowledgments}
This work was supported by NIH grants R01-HL138306 and  R01-ES016626.
{\it Conflict of Interest}: None declared.

\bibliographystyle{spmpsci}
\bibliography{proflik}

\end{document}